\documentclass[journal]{IEEEtran}

\usepackage{amsopn}
\usepackage{times}
\usepackage{graphicx}
\usepackage{nopageno}
\usepackage{eso-pic}
\usepackage{amssymb}
\usepackage{amsmath,epsfig,subfigure}
\usepackage{enumerate}
\usepackage{color}

\usepackage[numbers]{natbib}
\usepackage{algorithm}
\usepackage[noend]{algpseudocode}

\usepackage{longtable}
\usepackage{makecell}

\usepackage{tikz}
\def\checkmark{\tikz\fill[scale=0.4](0,.35) -- (.25,0) -- (1,.7) -- (.25,.15) -- cycle;} 

\usepackage{graphicx}
\usepackage[justification=centering]{caption}
\renewcommand{\thetable}{\arabic{table}}

\IEEEoverridecommandlockouts
\begin{document}
\title{Sorting with GPUs: A Survey}
\author{Dmitri I. Arkhipov, Di Wu, Keqin Li, and Amelia C. Regan
\thanks{D. I. Arkhipov and A. C. Regan are with the Department of Computer Science, University of California, Irvine, CA 92697, USA.} %
\thanks{D. Wu is with the Department of Computer Engineering, Hunan University, Changsha, Hunan 410082, China.} %
\thanks{K. Li is with the Department of Computer Science, State University of New York, New Paltz, New York 12561, USA.} %
}

\date{}
\maketitle

\begin{abstract}
Sorting is a fundamental operation in computer science and is a bottleneck in many important fields. Sorting is critical to database applications, online search and indexing, biomedical computing, and many other applications. The explosive growth in computational power and availability of GPU coprocessors has allowed sort operations on GPUs to be done much faster than any equivalently priced CPU. Current trends in GPU computing shows that this explosive growth in GPU capabilities is likely to continue for some time. As such, there is a need to develop algorithms to effectively harness the power of GPUs for crucial applications such as sorting.  
\end{abstract}


\section{Introduction}
In this survey, we explore the problem of sorting on GPUs. As GPUs allow for effective hardware level parallelism, research work in the area concentrates on parallel algorithms that are effective at optimizing calculations per input, internal communication, and memory utilization. Research papers in the area concentrate on implementing sorting networks in GPU hardware, or using data parallel primitives to implement sorting algorithms. These concentrations result in a partition of the body of work into the following types of sorting algorithms: parallel radix sort, sample sort, and hybrid approaches. 

The goal of this work is to provide a synopsis and summary of the work of other researchers on this topic, to identify and highlight the major avenues of active research in the area, and finally to highlight commonalities that emerge in many of the works we encountered. Rather than give complete descriptions of the papers referenced, we strive to provide short but informative summaries. This review paper is intended to serve as a detailed map into the body of research surrounding this important topic at the time of writing. We hope also to inform interested readers and perhaps to spark the interest of graduate students and computer science researchers in other fields. Complementary and in-depth discussions of many of the topics mentioned in this paper can be found in Bandyopadhyay et al.'s \cite{bandyopadhyay2013sorting} and in Capannini et al.'s \cite{capannini2012sorting} excellent review papers.

It appears that the main bottleneck for sorting algorithms on GPUs involves memory access. In particular, the long latency, limited bandwidth, and number of memory contentions (i.e., access conflicts) seem to be major factors that influence sorting time. Much of the work in optimizing sort on GPUs is centred around optimal use of memory resources, and even seemingly small algorithmic reductions in contention or increases in on chip memory utilization net large performance increases. 

Our key findings are the following:
\begin{itemize}
	\item Effective parallel sorting algorithms must use the faster access on-chip memory as much and as often as possible as a substitute to global memory operations.  
	\item Algorithmic improvements that used on-chip memory and made threads work more evenly seemed to be more effective than those that simply encoded sorts as primitive GPU operations.
	\item Communication and synchronization should be done at points specified by the hardware.
	\item Which GPU primitives (scan and 1-bit scatter in particular) are used makes a big difference. Some primitive implementations were simply more efficient than others, and some exhibit a greater degree of fine grained parallelism than others.
	\item A combination of radix sort, a bucketization scheme, and a sorting network per scalar processor seems to be the combination that achieves the best results.
	\item Finally, more so than any of the other points above, using on-chip memory and registers as effectively as possible is key to an effective GPU sort.
\end{itemize}	

We will first give a brief summary of the work done on this problem, followed by a review of the GPU architecture, and then some details on primitives used to implement sorting in GPUs. Afterwards, we continue with a more detailed review of the sorting algorithms we encountered, and conclude with a section examining aspects common to many of the papers we reviewed. 

\section{Retrospective} 
In \cite{batcher1968sorting, dowd1989periodic} Batcher’s bitonic and odd-even merge sorts are introduced, as well as Dowd’s periodic balanced sorting networks which were based on Batcher’s ideas. Most algorithms that run on GPUs use one of these networks to perform a per multiprocessor or per scalar processor sort as a component of a hybrid sort. Some of the earlier work is a pure mapping of one of these sorting networks to the stream/kernel programming paradigm used to program current GPU architectures.

Blelloch et al., and Dusseau et al., \cite{blelloch1991comparison,dusseau1996fast} showed that bitonic sort is often the fastest sort for small sequences but its performance suffers for larger inputs. This is due to the $O( n \log ^2 (n) )$ data access cost, which can result in high communication costs when higher level memory storage is required. In \cite{cederman2008practical, cederman2009sorting} Cederman et al. have adapted quick sort for GPUs, in Bandyopadhyay's adaptation \cite{bandyopadhyay2010grs} the researchers first partition the sequence to be sorted into sub-sequences, then sorts these sub-sequences and merges the sorted sub-sequences in parallel. 

The fastest GPU merge sort algorithm known at this time is presented by Davidson et al. \cite{davidson2012efficient} a title attributed to the greater use of register communications as compared to shared memory communication in their sort relative to other comparison based parallel sort implementations. 
Prior to Davidson's work, warp sort, developed by Ye et al. \cite{ye2010high} was the fastest comparison based sort, Ye et al. also attribute this speed to gains made through reducing communications delays by virtue of greater use of GPU privimitives, though in the case of warpsort, communication was handled by synchronization free inter-warp communication. Sample sort \cite{dehne2012deterministic} is reported to be about $30\%$ faster on average than the merge sort of \cite{satish2009designing} when the keys are 32-bit integers. This makes sample sort competitive with warp sort for 32-bit keys and for 64-bit keys, sample sort \cite{dehne2012deterministic} is on average twice as fast as the merge sort of \cite{satish2009designing}. Govindaraju et al. \cite{govindaraju2006gputerasort} discussed the question of sorting multi-terabyte data through an efficient external sort based on a hybrid radix-bitonic sort. The authors were able to achieve near peak IO performance by a novel mapping of the bitonic sorting network to GPU primitive operations (in the sense that GPU addressing is based off of pixel, texture and quadrilateral addressing and the network is mapped to such structures). Much of the work presented in the paper generalizes to any external scan regardless of the in-memory scan used in phase 1 of the external sort. 

Baraglia et al. investigate optimal block-kernel mappings of a bitonic network to the GPU stream/kernel architecture \cite{baraglia2009sorting}, their pure bitonic sort was able to beat \cite{cederman2008practical,cederman2009sorting} the quicksort of Cederman et al. for any number of record comparisons. However Manca et al.'s \cite{manca2015cuda} CUDA-quicksort which adapts Cederman's work to more optimally access GPU memory outperforms Baraglia et al.'s earlier work.    
Satish et al.'s 2009 \cite{satish2009designing} adaptation of radix sort to GPUs uses the radix 2 (i.e., each phase sorts on a bit of the key using the 1-bit scan primitive) and uses the parallel bitsplit technique. Le et al \cite{le2007broad} reduce the number of phases and hence the number of expensive scatters to global memory by using a larger radix, $2^b$ , for $b > 0$. The sort in each phase is done via a parallel count-sort implemented through a scan. Satish et al.\cite{satish2009designing} further improve the $2^b$ -radix sort by sorting blocks of data in shared memory before writing to global memory. For sorting $t$ values on chip with a t-thread block, In \cite{satish2009designing} Satish et al. found that the Batcher sorting networks to be substantially faster than either radix sort or \cite{cederman2008practical,cederman2009sorting}’s quicksort. Sundar et al.'s HykSort \cite{sundar2013hyksort} applied a splitter based sort more commonly used in sample sorts to develop a k-way recursive quicksort able to outperform comparable sample sort and bitonic sort implementations. Shamoto et al. \cite{shamoto2014large} HykSort with a theoretical any empirical analysis of the bottlenecks arising from different work partitions between splitters and local sorts as they arise in different hardware configurations. Beliakov et al. \cite{beliakov2015parallel} approach the problem of optimally determining splitters from an optimization perspective, formulating each splitter selection as a selection problem. Beliakov et al. are able to develop a competitive parallel radix sort based on their observation.      

The results of Leischner et al. and Ye et al. \cite{leischner2010gpu, ye2010high} indicated that the radix sort algorithm of \cite{satish2009designing} outperforms both warp sort \cite{ye2010high} and sample sort \cite{ye2014gpumemsort}, so the radix sort of \cite{satish2009designing} is the fastest GPU sort algorithm for 32-bit integer keys.

Chen et al. \cite{chen2009fast} expanded on the ideas of \cite{satish2009designing} to produce a hybrid sort that is able to sort not only integers as \cite{satish2009designing} can, but also floats and structs. Since the sort from \cite{satish2009designing} is a radix based sort the implementation is not trivially expandable to floats and structs, \cite{chen2009fast} produces an alternate bucketization scheme to the radix sort (the complexity of which is based on the bits in the binary representation of the input and explodes for floats). Chen et al. \cite{chen2009fast} extended the performance gains of \cite{satish2009designing} to datatypes other than ints. More recently \cite{satish2010fast, bandyopadhyay2010grs, merrill2011high} introduced significant optimizations to the work of \cite{chen2009fast}. Each in turn claiming the title of 'fastest GPU sort' by virtue of their optimisations. While not a competitive parallel sort in the sense of \cite{chen2009fast} and others mentioned herein, \cite{sun2009count} presents a detailed exposition of parallelizing count sort via GPU primitives. Count sort itself is used as a primitive subroutine in many of the papers presented here. Some recent works by Jan et al.
\cite{jan2013parallel,jan2012fast} implementing parallel butterfly sort on a variety of GPUs finds that the butterfly sorting network outperforms a baseline implementation of bitonic sort, odd-even sort, and rank sort. Ajdari et al. \cite{ajdari2015version} introduce a generalization of odd-even sorting applicable element blocks rather than elements so-as to better make use of the GPU data bus.

Tanasic et al. \cite{tanasic2013comparison} attack the problem of improved shared memory utilization from a novel perspective and develop a parallel merge sort focused on developing lighter merge phases in systems with multiple GPUs. Tanasic et al. concentrate on choosing pivots to guarantee even distribution of elements among the distinct GPUs and they develop a peer to peer algorithm for PCIe communication between GPUs to allow this desirable pivot selection. Works like \cite{jan2013parallel,jan2012fast} represent an important vector of research in examining  the empirical performance of classic sorting networks as compared to the more popular line of research in incremental sort performance improvement achieved by using non-obvious primitive GPU operations and improving cache coherence and shared memory use.     

\section{GPU Architecture}
\begin{figure*}
    \centering
    \includegraphics[width=0.9\textwidth]{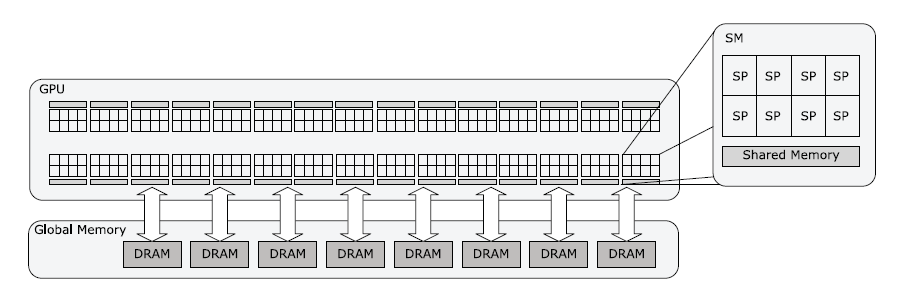}
    \caption{Leischner et al., 2010 \cite{leischner2010gpu} }
    \label{fig:Leischner}
\end{figure*}

\subsubsection*{Stream/Kernel Programming Model}
Nvida’s CUDA is based around the stream processing paradigm. Baraglia et al. \cite{baraglia2009sorting} give a good overview of this paradigm. A stream is a sequence of similar data elements, where the elements must share a common access pattern (i.e., the same instruction where the elements are different operands). A kernel is an ordered sequence of operations that will be performed on the elements of an input stream in the order that the elements are arranged in the stream. A kernel will execute on each individual element of the input stream. Kernels append the results of their operations on an output stream. Kernels are composed of parallelizable instructions. Kernels maintain a memory space, and must use that memory space for storing and retrieving intermediate results. Kernels may execute on elements in the input stream concurrently. Reading and operating on elements only once from memory, and using a large number of arithmetic operations per element are ideal ways for an implementation to be efficient. 

\subsubsection*{GPU Architecture}
The GPU consists of an array of streaming multiprocessors (SMs), each of which is capable of supporting up to 1024 co-resident concurrent threads. A single SM contains 8 scalar processors (SPs), each with 1024 32-bit registers. Each SM is also equipped with a 16KB on-chip memory, referred to as shared memory, that has very low access latency and high bandwidth, similar to an L1 cache.
The constant values from the above description are given in \cite{satish2009designing}, more recent hardware may have improved constant values.  
Besides registers and shared memory, on-chip memory shared by the cores in an SM also include constant and texture caches. 

\begin{figure}[htbp]
    \centering
    \includegraphics[width=0.40\textwidth]{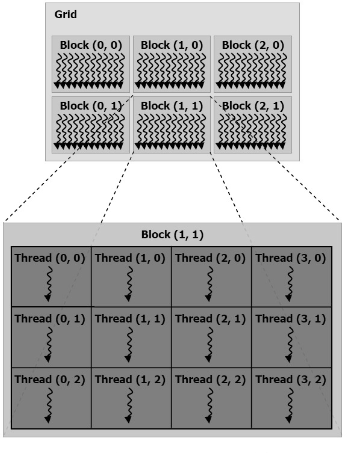}
    \caption{Nvidia, 2010 \cite{nvidia2010guide} }
    \label{fig:Bandy}
\end{figure}

A host program is composed of parallel and sequential components. Parallel components, referred to as kernels, may execute on a parallel device (an SM). Typically, the host program executes on the CPU and the parallel kernels execute on the GPU. A kernel is a SPMD (single program multiple data) computation, executing a scalar sequential program across a set of parallel threads. It is the programmer's responsibility to organize these threads into thread blocks in an effective way. 
Threads are grouped into a blocks that are operated on by a kernel. A set of thread blocks (referred to as a grid) is executed on an SM. An SP (also referred to as a core) within an SM is assigned multiple threads. As Satish et al. describe \cite{satish2009designing} a thread block is a group of concurrent threads that can cooperate among themselves through barrier synchronization and a per-block shared memory space private to that block. When invoking a kernel, the programmer specifies both the number of blocks and the number of threads per block to be created when launching the kernel. 
Communication between blocks of threads is not allowed and each must execute independently. 
Figure \ref{fig:Bandy} presents a visualization of grids, threads-blocks, and threads.

\section{GPU Architecture}

Threads are executed in groups of 32 called warps. \cite{satish2009designing}. 
The number of warps per block is also configurable and thus determined by the programmer, however in \cite{merrill2011high} and likely a common occurrence, each SM contains only enough ALUs (arithmetic logic units) to actively execute one or two warps. 
Each thread of an active warp executes on one of the scalar processors (SPs) within a single SM. 
In addition to SIMD and vector processing capabilities (which exploit the data parallelism in the executed program), each SM can also exploit instruction-level parallelism by evaluating multiple independent instructions simultaneously with different ALUs. In \cite{brodtkorb2013graphics} task level parallelism is achieved by executing multiple kernels concurrently on different SMs. 
The GPU relies on multi-threading rather than caching to hide the latency of external memory transactions. The execution state of a thread is usually saved in the register file.	

\subsubsection*{Synchronization}
All thread management, including creation, scheduling, and barrier synchronization is performed entirely in hardware by the SM with essentially zero overhead. 
Each SM swaps between warps in order to mask memory latencies as well as those from pipeline hazards. This translates into tens of warp contexts per core, and tens-of-thousands of thread contexts per GPU microprocessor. However, by increasing the number of threads per SM, the number of available registers per thread is reduced as a result. 
Only threads (within a thread block) concurrently running on the same SM can be synchronized, therefore in an efficient algorithm, synchronization between SMs should not be necessary or at least not be frequent. 
Warps are inherently synchronized at a hardware level. 

Once a block begins to execute on an SM, the SM executes it to completion, and it cannot be halted. When an SM is ready to execute the next instruction, it selects a warp that is ready (i.e., its threads are not waiting for a memory transaction to complete) and executes the next instruction of every thread in the selected warp. Common instructions are executed in parallel using the SPs (scalar processors) in the SM. Non-common instructions are serialized. Divergences are non-common instructions resulting in threads within a warp following different execution paths. To achieve peak efficiency, kernels should avoid execution divergence. As noted in \cite{satish2009designing} divergence between warps, however, introduces no performance penalty.

\subsubsection*{Memory}
Each thread has access to private registers and local memory. Shared memory is accessible to all threads of a thread block. And global memory is accessible by all threads and the host code. All the memory types mentioned above are random access read/write memory. A small amount of global memory, constant and texture memory is read-only; this memory is cached and may be accessed globally at high speed.
The latency of working in on-chip memory is roughly 100 times faster than in external DRAM, thus the private on-chip memory should be fully utilised, and operations on global memory should be reduced. 
As noted by Brodtkorb et al. \cite{brodtkorb2013graphics} GPUs have L1 and L2 caches that are similar to CPU caches (i.e., set-associativity is used and blocks of data are transferred per I/O request). Each L1 cache is unique to each SM, whereas the L2 cache is shared by all SMs. The L2 cache can be disabled during compile time, which allows transference of data less than a full cache line.
Only one thread can access a particular bank (i.e., a contiguous set of memory) in shared memory at a time. As such, bank conflicts can occur whenever two or more threads try to access the same bank. Typically, shared memory (usually about 48KB) consists of 32 banks in total. A common strategy for dealing with bank conflicts is to pad the data in shared memory so that data which will be accessed by different threads are in different banks.
When dealing with global memory in the GPU, cache lines (i.e., contiguous data in memory) are transferred per request, and even when less data is sent, the same amount of bandwidth as a full cache line is used, where typically a full cache line is at least 128 bytes (or 32 bytes for non-cached loads). As such, data should be aligned according to this length (e.g., via padding). It should also be noted that the GPU uses memory parallelism, where the GPU will buffer memory requests in an attempt to fully utilize the memory bus, which can increase the memory latency.

The threads of a warp are free to load from and store to any valid address, thus supporting general gather and scatter access to memory. However, in \cite{he2007efficient} He et al. showed that sequential access to global memory takes ~5 ms, while random access takes ~177 ms. An optimization that takes advantage of this fact involves a multi-pass scheme where contiguous data is written to global memory in each pass rather than a single pass scheme where random access is required.
When threads of a warp access consecutive words in memory, the hardware is able to coalesce these accesses into aggregate transactions within the memory system, resulting in substantially higher memory throughput. Peters et al. \cite{peters2011fast} claimed that a half-warp can access global memory in one transaction if the data are all within the same global memory segment. Bandyopadhyay et al. \cite{bandyopadhyay2011sorting} showed that using memory coalescing progressively during the sort with the By Field layout (for data records) achieves the best result when operating on multi-field records. Govindaraju et al. \cite{govindaraju2006gputerasort} showed that GPUs suffer from memory stalls, both because the memory bandwidth is inadequate and lacks a data-stream approach to data access.
The CPU and GPU both simultaneously send data to each other as long as there are no memory conflicts. Furthermore, it is possible to disable the CPU from paging memory (i.e., writing virtual memory to disk) in order to keep the data in main memory. Clearly, this can improve access time, since the data is guaranteed to be in main memory. It is also possible to use write-combining allocation. That is, the specified area of main memory is not cached. This is useful when the area is only used for writing. Additionally, ECC (Error Correcting Code) can be disabled to better utilize bandwidth and increase available memory, by removing the extra bits used by ECC.

\section{GPU Primitives}
Certain primitive operations can be very effectively implemented in parallel, particularly on the architecture of the GPU. Some of these operations are used extensively in the sort implementation we encountered. The two most commonly used primitives were prefix sum, and 1-bit scatter.  

\subsubsection*{Scan/Prefix Sum}
The prefix sum (also called scan) operation is widely used in the fastest parallel sorting algorithms. Segupta et al. \cite{sengupta2007scan} showed an implementation of prefix sum which consists of $2 \log (n)$ step phases, called reduce and downsweep. In another paper, Segupta et al. \cite{sengupta2008efficient} revealed an implementation which consists of only $\log (n)$ steps, but is more computationally expensive.

\subsubsection*{1-Bit Scatter}
Satish et al. \cite{satish2010fast} explained how the 1-bit scatter, a split primitive, distributes the elements of a list according to their category. For example, if there is a vector of $N$ 32-bit integers and a 1-bit split operation is performed on the integers, and the 30th bit is chosen as the split bit, then all elements in the vector that have a 1 in the 30th bit will move to the upper half of the vector, and all other elements (which contain a 0) will move to the lower half of the vector.  This distribution of values based on the value of the bit at a particular index is characteristic of the 1-bit split operation.

\section{GPU Trends}
In this section, we mention some trends in GPU development.
GPUs have become popular among the research community as its computational performance has developed well above that of CPUs through the use of data parallelism. As of 2007, \cite{govindaraju2006gputerasort} states that GPUs offer $10X$ more memory bandwidth and processing power than CPUs; and this gap has increases since, and this increase is continuing, the computational performance of GPUs is increasing at a rate of about twice a year.
Keckler et al. \cite{keckler2011gpus} compared the rate of growth between memory bandwidth and floating point performance of GPUs. The figure below from that article helps illustrate this point.

\begin{figure}[htbp]
    \centering
    \includegraphics[width=0.48\textwidth]{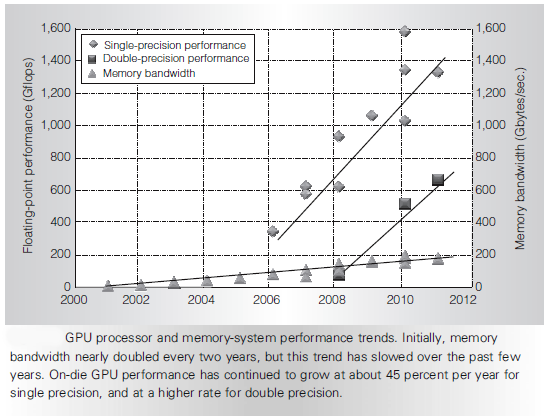}
    \caption{Keckler et al., 2011 \cite{keckler2011gpus} }
    \label{fig:Keckler}
\end{figure}

Floating point performance is increasing at a much faster rate than memory bandwidth (a factor of 8 times more quickly) and Nvidia expects this trend to continue through 2017. As such, it is expected that memory bandwidth will become a source of bottleneck for many algorithms. It should be noted that latency to and from memory is also a major bottleneck in computing. Furthermore, the trend in computing architectures is to continue reducing power consumption and heat while increasing computational performance by making use of parallelism, including increasing the number of processing cores.

\section{Cluster And Grid GPU Sorting}
An important direction for research in this subject is the elegant use of heterogeneous grids/clusters/clouds of computers with banks of GPUs while sorting. In their research White et al.\cite{white2012cuda} take a first step in this direction by implementing a 2-phase protocol to leverage the well known parallel bitonic sort algorithm in a cluster environment with heterogeneous compute nodes. In their work data was communicated through the MPI message passing standard. This brief work, while a straight forward adaptation of standard techniques to a cluster environment gives important empirical benchmarks for the speed gains available when sorting in GPU clusters and the data sizes necessary to access these gains. Shamoto et al. \cite{shamoto2014large} give benchmarks representative of more sophisticated grid based GPU sort. Specifically they describe where bottlenecks arise under different hardware environments using Sundar et al.'s Hyksort \cite{sundar2013hyksort} (a splitter based GPU-quicksort) as an experimental baseline. Shamoto et al. analyze the splitter based approach to GPU sorting giving the hardware and algorithmic conditions under which the communications delays between different network components dominate the search time, and the conditions at which local sort dominates. Zhong et al. approach the problem from a different perspective in their paper \cite{zhong2012efficient} the researchers improve sorting throughput by introducing a scheduling strategy to the cluster computation which ensures that each compute node recives the next required data block before it completes calculations on its current block.
      
\section*{Sorting Algorithms}
Next we describe the dominant techniques for GPU based sorting.  We found that much of the work on this topic focuses on a small number of sorting algorithms, to which the authors have added some tweaks and personal touches. However, there are also several non-standard algorithms that a few research teams have discovered which show impressive results. Generally, sorting networks such as the bitonic sorting network appear to be the algorithms that have obtained the most attention among the research community with many variations and applications in the papers that we have read. However, radix sort, quicksort and sample sort are also quite popular. Many of the algorithms serve distinct purposes such that some algorithms performs better than others depending on the specified tasks. In this section we briefly mention some of the most popular sorting algorithms, along with their performance and the contributions of the authors.

\subsubsection*{Sorting Networks}
Effective sorting algorithms for GPUs have tended to implement sorting networks as a component. In the following sections we review the mapping of sorting networks to the GPU as a component of current comparison based sorts on GPUs.
Dowd et al. \cite{dowd1989periodic} introduced a lexicon for discussing sorting networks that is used in the sections below. A sorting network is comprised of a set of two-input, two-output comparators interconnected in such a way that, a short time after an unordered set of items are placed on the input ports, they will appear on the output ports with the smallest value on the first port, the second smallest on the second port, etc. The time required for the values to appear at the output ports, that is, the sorting time, is usually determined by the number of phases of the network, where a phase is a set of comparators that are all active at the same time. The unordered set of items are input to the first phase; the output of the $i$'th phase becomes the input to the $i + 1$th phase. A phase of the network compares and possibly rearranged pairs of elements of x. It is often useful to group a sequence of connected phases together into a block or stage.  A periodic sorting network is defined as a network composed of a sequence of identical blocks.

\section*{Comparison-Based Sorting}

\subsubsection*{Bitonic Sort}
Satish et al. \cite{satish2009designing} produced a landmark bitonic sort implementation that is referred to, modified, and improved in other papers, therefore it is presented in some detail.  The merge sort procedure consists of three steps: 

\begin{enumerate}
	\item Divide the input into $p$ equal-sized tiles.
	\item Sort all $p$ tiles in parallel with $p$ thread blocks.
	\item Merge all $p$ sorted tiles.
\end{enumerate}

The researchers used Batcher's odd-even merge sort in step 2, rather than bitonic sort, because their experiments showed that it is roughly $5-10\%$ faster in practice. The procedure spends a majority of the time in the merging process of Step (3), which is accomplished with a pair-wise merge tree of $\log (p)$ depth. Each level of the merge tree pairs corresponding odd and even sub-sequences. During phase 3 they exploit parallelism by computing the final position of elements in the merged sequence via parallel binary searches. They also used a novel parallel merging technique. As an example, suppose we want to merge sequences $A$ and $B$, such that $C = merge(A,B)$. Define $rank(e, C)$ of an element $e$ to be its numbered position after $C$ is sorted. If the sequences are small enough then we can merge them using a single thread block. For the merge operation, since both $A$ and $B$ are already sorted, we have $rank(a_i , C) = i + rank(a_i , B)$, where $rank(a_i , B)$ is simply the number of elements $b_j \in B$ such that $b_j < a_i$ which can be computed efficiently via binary search. 
Similarly, the rank of the elements in $B$ can be computed. 
Therefore, we can efficiently merge these two sequences by having each thread of the block compute the rank of the corresponding elements of $A$ and $B$ in $C$, and subsequently writing those elements to the correct position. Since this can be done in on-chip memory, it will be very efficient. In \cite{satish2009designing} Satish et al. merge of larger arrays are done by dividing the arrays into tiles of size at most $t$ that can be merged independently using the block-wise process. To do so in parallel, begin by constructing two sequences of splitters $S_A$ and $S_B$ by selecting every $t$-th element of A and B, respectively.
By construction, these splitters partition $A$ and $B$ into contiguous tiles of at most $t$ elements each. Construct a merged splitter set $S = merge(S_A , S_B )$, which we achieve with a nested invocation of our merge procedure. Use the combined set $S$ to split both $A$ and $B$ into contiguous tiles of at most $t$ elements. To do so, compute the rank of each splitter $s$ in both input sequences. The rank of a splitter $s = a_i$ drawn from $A$ is obviously $i$, the position from which it was drawn. To compute its rank in $B$, first compute its rank in $S_B$. Compute this directly as $rank(s, S_B ) = rank(s, S) - rank(s, S_A )$, since both terms on the right hand side are its known positions in the arrays $S$ and $S_A$.  Given the rank of $s$ in $S_B$ , we have now established a t-element window in $B$ in which $s$ would fall, bounded by the splitters in $S_B$ that bracket $s$. Determine the rank of $s$ in $B$ via binary search within this window. 
Blelloch et al. and Dusseau et al. showed \cite{blelloch1991comparison, dusseau1996fast} that bitonic sort works well for inputs of a small size, Ye et al. \cite{ye2010high} identified a way to sort a large input in small chunks, where each chunk is sorted by a bitonic network and then the chunks are merged by subsequent bitonic sorts. During the sort warps progress without any explicit synchronization, and the elements being sorted by each small bitonic network are always in distinct memory location thereby removing any contention issues. Two sorted tiles are merged as follows: Given tile $A$ and tile $B$ compose two tiles $L(A)L(B)$ and $H(A)H(B)$ where $L(X)$ takes the smaller half of the elements in $X$, and $H(X)$ the larger half. $L(A)L(B)$ and $H(A)H(B)$ are both bitonic sequences and can be sorted by the tile-sized bitonic sorting networks of warpsort. Several hybrid algorithms such as Chen et al.'s \cite{chen2009fast} perform an in memory bitonic sort per bucket after bucketizing the input elements. Braglia et al. \cite{baraglia2009sorting} discussed a scheme for mapping contiguous portions of Batcher's bitonic sorting network to kernels in a program via a parallel bitonic sort within a GPU. A bitonic sort is composed of several phases. All of the phases are fitted into a sequence of independent blocks, where each of the blocks correspond to a separate kernel. Since each kernel invocation implies I/O operations, the mapping to a sequence of blocks should reduce the number of separate I/O transmissions. It should be noted that communication overhead is generated whenever the SM processor begins the execution of a new stream element. When this occurs the processor flushes the results contained in its shared memory, and then fetches new data from the off-chip memory. We must establish the number of consecutive steps to be executed per kernel. In order to maintain independence of elements, the number of elements received by the kernel doubles every time we increase the number of steps by one. So, the number of steps a kernel can cover is bounded by the number of items that is possible to be included in the stream element. Furthermore, the number of items is bounded by the size of the shared memory available for each SIMD processor. If each SM has 16 KB of local memory, then we can specify a partition consisting of $SH = 4K$ items, for 32-bit items. $(32bits \times 4K = 15.6KB)$ Moreover such a partition is able to cover ``at least" $sh = lg(SH) = 12$ steps (because we assume items within the kernel are only compared once). If a partition representing an element of the stream contains SH items, and the array to sort contains $N = 2^n$ items, then the stream contains $b = N/SH = 2^{n-sh}$ elements. 
The first kernel can compute $\frac{(sh)\times(sh+1)}{2}$ steps because our conservative assumption is that each number is only compared to one other number in the element. In fact, bitonic sort is structured so that many local comparisons happen at the start of the algorithm. This optimal mapping is illustrated below:

\begin{figure*}
    \centering
    \includegraphics[width=0.9\textwidth]{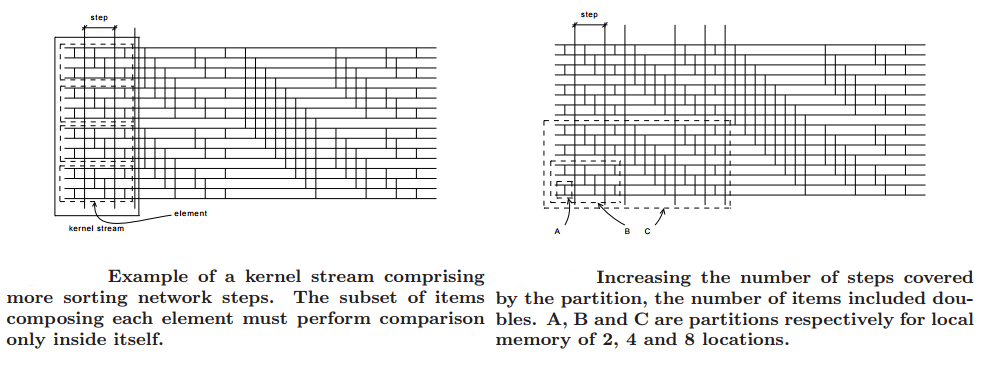}
    \caption{Baraglia et al., 2009 \cite{baraglia2009sorting} }
    \label{fig:Baraglia}
\end{figure*}

Satish et al. \cite{satish2010fast} performed a similar though better implementation of bitonic sort as \cite{satish2009designing}, they found that bitonic merge sort suffers from the overhead of index calculation and array reversal, and that  a radix sort implementation is superior to merge sort as there are less calculations performed overall. In \cite{govindaraju2006gputerasort} Govindaraju et al. explored mapping a bitonic sorting network to operations optimized on GPUs (pixel and quadrilateral computations on 4 channel arrays). The work explored indexing schemes to capitalize on the native 2D addressing of textures and quadrilaterals in the GPU. In a GPU, each bitonic sort step corresponds to mapping values from one chunk in the input texture to another chunk in the input texture using the GPU’s texture mapping hardware. 

The initial step of the texturing process works as follows. First, a 2D array of data values representing the texture is specified and transmitted to the GPU. Then, a 2D quadrilateral is specified with texture coordinates at the appropriate vertices. For every pixel in the 2D quadrilateral, the texturing hardware performs a bilinear interpolation of the lookup coordinates at the vertices. The interpolated coordinate is used to perform a 2D array lookup by the MP. This results in the larger and smaller values being written to the higher and lower target pixels. As GPUs are primarily optimized for 2D arrays, they map the 1D array of data onto a 2D array. The resulting data chunks are 2D data chunks that are either row-aligned or column-aligned. Govindaraju et al. \cite{govindaraju2006gputerasort} used the following texture representation for data elements. Each texture is represented as a stretched 2D array where each texel holds 4 data items via the 4 color channels. Govindaraju et al. \cite{govindaraju2006gputerasort} found that this representation is superior to earlier texture mapping schemes such as the scheme used in GPUSort, since more intra-chip memory locality is utilised leading to fewer global memory access.  

Peters et al. \cite{peters2011fast} implemented an in-place bitonic sorting algorithm, which was asserted to be the fastest comparison-based algorithm at the time (2011) for approximately less than $2^{26}$ elements, beating the bitonic merge sort of \cite{satish2009designing}. An important optimization discovered in \cite{peters2011fast} involves assigning a partition of independent data used across multiple consecutive steps of a phase in the bitonic sorting algorithm to a single thread. However, the number of global memory access is still  and as such this algorithm can not be faster than algorithms with lower order of global memory access (such as sample sort) when there is a sufficient number of elements to sort.

The Adaptive Bitonic Sort (ABS) of Greb et al. \cite{greb2006gpu} can run in $O(\log ^2 (n))$ time and is based on Batcher's bitonic sort. ABS belongs to the class of algorithms known as \textit{adaptive bitonic algorithms}; a class introduce and analyzed in detail Bilardi et al's \cite{bilardi1989adaptive}. In ABS a pivot point in the calculations for bitonic sort is found such that if the first $j$ points of a set $p$ and the last $j$ points in a set $q$ are exchanged, the bitonic sequences $p'$ and $q'$ are formed, which are the resulting Low and High bitonic sequences at the end of the step.  However, Greb et al. claimed that $j$ may be found by using a binary search. They utilize a binary tree in the sorting algorithm for storing the values into the tree in so as to find $j$ in $\log (n)$ time.  This also gives the benefit that entire sub trees can be replaced using simple pointer exchange, which greatly helps find $p'$ and $q'$.

The whole process requires $\log (n)$ comparisons and less than $2 \log (n) $ exchanges.  Using this method gives a recursive merge algorithm in $O(n)$ time which is called adaptive bitonic merge. In this method the tree doesn't have to rebuild itself at each stage. In the end, this results in the classic sequential version of adaptive bitonic sorting which runs in $O(n \log (n))$ total time, a significant improvement over both Batcher’s bitonic sorting network and Dowd’s periodic sorting network \cite{batcher1968sorting, dowd1989periodic}. For the stream version of the algorithm, as with the sequential version, there is a binary tree that is storing the values of the bitonic sort.  There are however $2^k$ subtrees for each pair of bitonic trees that is to be merged, which can all be compared in parallel. Zachmann gives an excellent and modern review of the theory and history of both bitonic sorting networks and adaptive bitonic sorting in \cite{zachmann2013adaptive}.

A problem with implementing the algorithm in the GPU architecture however is that the stream writes which must be done sequentially since GPUs (as of the writing of \cite{greb2006gpu}) are inefficient at doing random write accesses. But since trees use links from children to parents, the links can be updated dynamically and their future positions can be calculated ahead of time eliminating this problem. However, GPUs may not be able to overwrite certain links since other instances of binary trees might still need them, which requires us to append the changes to a stream that will later make the changes in memory.  
This brings up the issue of memory usage.  In order to conserve memory, they set up a storage of $n/2$ node pairs in memory.  At each iteration $k$, when $k = 0$, all levels $0$ and below are written and no longer need to be modified. At $k = 1$, all levels 1 and below are written and therefore no longer need to be modified and so on. This allows the system to maintain a constant amount of memory in order to make changes.

Since each stage $k$ of a recursion level $j$ of the adaptive bitonic sort consists of $j - k$ phases, $O( \log (n) )$ stream operations are required for each stage. Together, all $j$ stages of recursion level $j$ consist of $\frac{j^2}{2} + \frac{j}{2}$ phases in total. Therefore, the sequential execution of these phases requires $O( \log ^2 (n) )$ stream operations per recursion level and, in total, $O( \log ^3 (n) )$ stream operations for the whole sort algorithm.  This allows the algorithm to achieve the optimal time complexity of $O(\frac{n \log (n)}{p})$ for up to $p = \frac{n}{ \log ^2 (n) }$ processors. Greb et al. claimed that they can improve this a bit further to $O( \log ^2 (n) )$ stream operations by showing that certain stages of the comparisons can be overlapped. Phase $i$ of the comparison $k$ can be done immediately after phase $i+1$ of comparison $k-1$, and by doing so, this overlap allows us  to execute these phases at every other step of the algorithm, allowing us $O( \log ^2 (n) )$ time. This last is a similar analysis to that of Baraglia et al. \cite{baraglia2009sorting}. Peters et al. \cite{peters2012novel} built upon adaptive bitonic sort and created a novel algorithm called Interval Based Rearrangement (IBR) Bitonic Sort. As with ABS, the pivot value $q$ must be found for the two sequences at the end of the sort that will result in the complete bitonic sequence, which can be found in $O( \log (M) )$ time where $M$ is the length of the bitonic sequence $E$. Prior work has used binary trees to find the value $q$, but Peters et al. used interval notation with pointers (offset x) and length of sub-sequences (length l).  Overall, this allows them to see speed-ups of up to $1.5$ times the fastest known bitonic sorts. Kipfer et al. \cite{kipfer2004uberflow} introduced a method based off of bitonic sort that is used for sorting 2D texture surfaces.  If the sorting is row based, then the same comparison that happens in the first column happens in every row. In the $k'th$ pass there are three facts to point out.

\begin{enumerate}
	\item The relative address or offset $(delta-r)$ of the element that has to be compared is constant.
	\item This offset changes sign every $2^{k-1}$ columns.
	\item Every $2^{k-1}$ columns the comparison operation changes as well.
\end{enumerate}	

The information needed in the comparator stages can thus be specified on a per-vertex basis by rendering column aligned quad-strips covering $2^k \times n$ pixels.  In the $k'th$ pass $\frac{n}{2^k}$ quads are rendered, each covering a set of $2^k$ columns. The constant offset ($\delta r$) is specified as a uniform parameter in the fragment program. The sign of this offset, which is either $+1$ or $-1$, is issued as varying per-vertex attribute in the first component $(r)$ of one of the texture coordinates. The comparison operation is issued in the second component $(s)$ of that texture coordinate. A less than comparison is indicated by $1$, a larger than comparison is $-1$. In a second texture coordinate the address of the current element is passed to the fragment program. The fragment program in pseudo code to perform the Bitonic sort is given in \cite{kipfer2004uberflow}, and replicated below:

\begin{algorithmic}
	\State $OP1\gets TEX1[r_{2}, s_{2}]$
	\If {$r_{1} < 0$}
    		\State $sign = -1$
    	\Else
    		\State  $sign = +1$
    	\EndIf
 	\State $OP2\gets TEX1[r_{2} + sign \times \delta r, s_{2}]$
 	\If {$OP1.x \times s_1 < OP2.x \times s_1$}
    		\State $output = OP_1$
    	\Else
    		\State $output = OP_2$
    	\EndIf 
\end{algorithmic}

Note that $OP1$ and $OP2$ refer to the operands of the comparison operation as retrieved from the texture $TEX1$. 
The element mapping in \cite{kipfer2004uberflow} is also discussed in \cite{govindaraju2006gputerasort}.

\subsubsection*{Periodic Balance Sorting Network}
Govindaraju et al. \cite{govindaraju2005fast} used the Periodic Balance Sorting Network of \cite{dowd1989periodic} as the basis of their sorting algorithm. This algorithm sorts an input of n elements in  steps. The output of each phase is used as the input to the subsequent phase. In each phase, the algorithm decomposes the input (a 2D texture) into blocks of equal sizes, and performs the same set of comparison operations on each block. If a block’s size is $B$, a data value at the i’th location in the block is compared against an element at the $(B - i)'th$ location in the block. If $i \geq B/2$ , the maximum of two elements being compared is placed at the i-th location, and if $i < B/2$ , the minimum is placed at the $i$'th location. The algorithm proceeds in $\log (n)$ stages, and during each stage $\log (n)$ phases are performed. 

At the beginning of the algorithm, the block size is set to $n$. At the end of each step, the block size is halved. At the end of $\log (n)$ stages, the input is sorted. In the case when $n$ is not a power of 2, $n$ is rounded to the next power of 2 greater than $n$. Based on the block size, the authors partition the input (implicitly) into several blocks and a Sortstep routine is called on all the blocks in the input. The output of the routine is copied back to the input texture and is used in the next step of the iteration. At the end of all the stages, the data is read back to the CPU. 

\begin{figure}[htbp]
    \centering
    \includegraphics[width=0.48\textwidth]{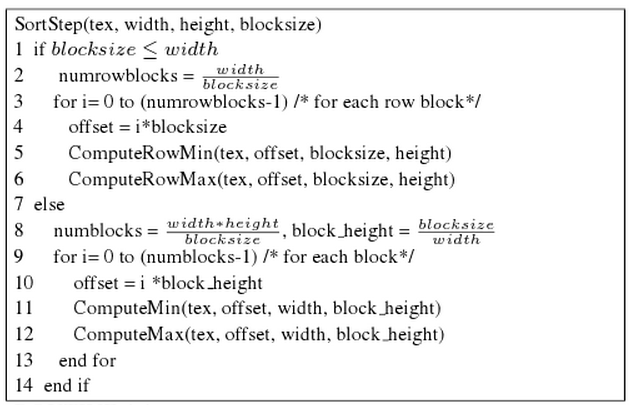}
    \caption{Govindaraju et al., 2005 \cite{govindaraju2005fast}}
    \label{fig:Govindaraju}
\end{figure}

The authors improve the performance of the algorithm by optimizing the performance of the Sortstep routine.  Given an input sequence of length $n$, they store the data values in each of the four color components of the 2D texture, and sort the four channels in parallel this is similar to the approach explored in \cite{govindaraju2006gputerasort} and in GPUSort. The sorted sequences of length $n/4$ are read back by the CPU and a merge operation is performed in software. The merge routine performs $O(n)$ comparisons and is very efficient.

Overall, the algorithm performs a total of $(n + n \log_2( n/4 ))$ comparisons to sort a sequence of length $n$.  In order to observe the $O( \log_2(n))$ behavior, an input size of $8M$ was used as the base reference for $n$ and estimated the time taken to sort the remaining data sizes.  The observations also indicate that the performance of the algorithm is around 3 times slower than optimized CPU-based quicksort for small values of n $(n < 16K)$. If the input size is small and the sorting time is slow because of it, the overhead cost of setting up the sorting function can dominate the overall running time of the algorithm.

\subsubsection*{Vector Merge Sort}
Sintorn et al. \cite{sintorn2008fast} introduced Vector-Merge Sort (VMS) which combines merge sort on four float vectors. While \cite{sintorn2008fast} and \cite{govindaraju2005fast} both deal with 2D textures, \cite{govindaraju2005fast} used a specialized sorting algorithm, while \cite{sintorn2008fast} employed VMS which much more closely resembles how a bitonic sort would work. A histogram is used to determine good pivot points for improving efficiency via the following steps:  

\begin{enumerate}
	\item Histogram of pivot points - find the pivot points for splitting the $N$ points into $L$ sub lists. $O(N)$ time.
	\item Bucket sort - split the list up into $L$ sub lists.  $O(N)$ 
	\item Vector-Merge Sort (VMS) - executes the sort in parallel on the $L$ sub lists.  $O(N \log (L) )$
\end{enumerate}	

\noindent To find the histogram of pivot points, there is a set of five steps to follow in the algorithm:

\begin{enumerate}
	\item The initial pivot points are chosen simply as a linear interpolation from the min value to the max value.
	\item The pivot points are first uploaded to each multiprocessor's local memory. One thread is created per element in the input list and each thread then picks one element and finds the appropriate bucket for that element by doing a binary search through the pivot points. When all elements are processed, the counters will be the number of elements that each bucket will contain, which can be used to find the offset in the output buffer for each sub-list, and the saved value will be each element's index into that sub list.
	\item Unless the original input list is uniformly distributed, the initial guess of pivot points is unlikely to result in a fair division of the list. However, using the result of the first count pass and making the assumption that all elements that contributed to one bucket are uniformly distributed over the range of that bucket, we can easily refine the guess for pivot points. Running the bucket sort pass again, with these new pivot points, will result in a better distribution of elements over buckets. If these pivot points still do not divide the list fairly, then they can be further refined. However, a single refinement of pivot points was usually sufficient.
	\item Since the first pass is usually just used to find an interpolation of the min and max, we simply ignore the first pass by not storing the bucket or bucket-index and instead just using the information to form the histogram. 
	\item Splitting the list into $d = 2 \times p$ sub-lists (where $p$ represents the number of processors) helps keep efficiency up and the higher number of sub-lists actually decreases the work of merge sort. However, the binary search for each bucket sort thread would take longer with more lists.
\end{enumerate}	

Polok et al. \cite{polok2014fast} give an improvement speeding up on register histogram counter accumulation mentioned in the algorithm above by using synchronization free warp-synchronous operations. Polok et al.'s optimization applies the techniques first highlighted in the context of Ye et al.'s  \cite{ye2010high} Warpsort to the context of comparison free sorting.  

In bucket sort, first we take the list of $L-1$ suggested pivot points that divide the list into L parts, then we count the number of elements that will end up in each bucket, while recording which bucket each element will end up in and what index it will have in that bucket. The CPU then evaluates if the pivot points were well chosen by looking at the number of elements in each bucket by how balanced the buckets are by their amount. Items can be moved between lists to make the lists roughly equal in size.

Starting VMS, the first step is to sort the lists using bitonic sort. The output is put into two arrays, $A$ and $B$, $a$ will be a vector taken from $A$ and $b$ from $B$, such that $a$ consists of the four lowest floats and $b$ is the four greatest floats.  These vectors are internally sorted. $a$ is then output as the next vector and $b$ takes its place. A new $b$ is found from $A$ or $B$ depending on which has the lowest value.  This is repeated recursively until $A$ and $B$ are empty.
Sintorn et al. found that for $1-8$ million entries, their algorithm was $6-14$ times faster than CPU quicksort. For $512k$ entries, their algorithm was slower than bitonic and radix sort.  However, those algorithms require powers of $2$ entries, while this one works on entries of arbitrary size.

\subsubsection*{Sample Sort}
Leischner et al. \cite{leischner2010gpu} presented the first implementation of a randomized sample sort on the GPU. In the algorithm, samples are obtained by randomly indexing into the data in memory. The samples are then sorted into a list. Every $K$’th sample is extracted from this list and used to set the boundaries for buckets. Then all of the n elements are visited to build a histogram for each bucket, including the tracking of the element count. Afterwards a prefix sum is performed over all of the buckets to determine the buckets’ offset in memory. Then the elements’ location in memory are computed and they are put in their proper position. Finally, each individual bucket is sorted, where the sorting algorithm used depends on the number of elements in the bucket. If there is too many elements within the bucket, then the above steps are repeated to reduce the number of elements per bucket. If the data size (within the bucket) is sufficiently small then quicksort is used, and if the data fits inside shared memory then Batcher’s odd-even merge sort is used. In this algorithm, the expected number of global memory access is $O(n \log_k(frac{n}{m}))$, where $n$ is the number of elements, $k$ is the number of buckets, and $M$ is the shared memory cache size.

Ye et al. \cite{ye2014gpumemsort} and Dehne et al. \cite{dehne2012deterministic} independently built on previous work on regular sampling by Shi et al. \cite{shi1992parallel}, and used it to create more efficient versions of sample sort called GPUMemSort and Deterministic GPU Sample Sort, respectively. Dehne et al. developed the more elegant algorithm of the two, which consists of an initial phase where the $n$ elements are equally partitioned into chunks, individually sorted, and then uniformly sampled. This guarantees that the maximum number of elements in each bucket is no greater than $\frac{2n}{s}$, where $s$ is the number of samples and $n$ is the total number of elements. A graph revealing the running time of the phases in Deterministic GPU Sample Sort is shown below. Both of the algorithms, which use regular sampling, achieve a consistent performance close to the best case performance of randomized sample sort, which is when the data has a uniform distribution. It should also be noted that the implementation of Deterministic GPU Sample Sort uses data in a more efficient manner than randomized sort, resulting in the ability to operate on a greater amount of data.

\begin{figure}[htbp]
    \centering
    \includegraphics[width=0.5\textwidth]{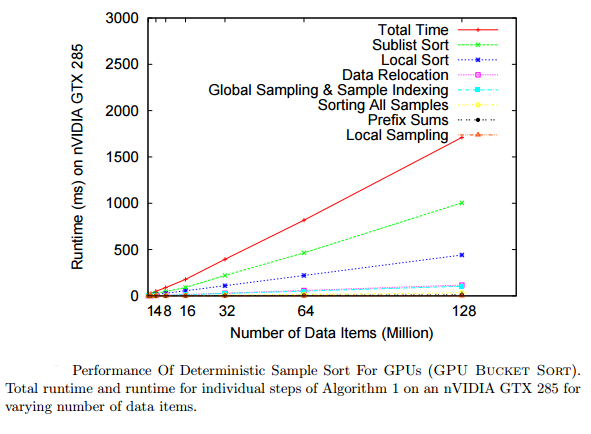}
    \caption{Dehne et al., 2010 \cite{dehne2012deterministic} }
    \label{fig:Dehne}
\end{figure}

\subsubsection*{Non-Comparison-Based}
The current record holder for fastest integer sort is one of the variations of radix sort.  Radix and radix-hybrid sorts are highly parallelizable and are not lower bounded by $O(n \log (n))$ time unlike sorting algorithms like quicksort. Since \cite{satish2009designing} radix-hybrid sorting algorithms have been the fastest among parallel GPU sorting algorithms, many implementations and hybrid algorithms focus on developing new variations of these sorting algorithms in order to take advantage of their parallel scalability and time bounds.

\section{Count Sort, Radix Sort, and Hybrids}

\subsection{Parallel Count Sort, A Component of Radix Sort}
Sun et al. \cite{sun2009count} introduced parallel count sort, based on classical counting sort. While it is not a competitive sort on its own, it serves as a primitive operation in most radix sort implementations. Count sort proceeds in 3 stages, firstly a count of the number of times each element occurs in the input sequence, then a prefix-sum on these counts, and finally a gather operation moving elements from the input array to the output at indices given by the values in the prefix-sum array. Sun et al. \cite{sun2009count} gave code for this algorithm. Each of the 3 stages are trivially parallelizable and require little communication, and synchronization, making count sort an excellent candidate for parallelization. Duplicate elements cause a huge increase in sort time, if there are no duplicates, the algorithm runs $100$ times faster, this is because duplicate elements are the only elements for which some inter-thread communication must occur. Satish et al.’s \cite{satish2009designing} radix sort Implementation set the standard for parallel integer sorts of the time, and in nearly every paper since reference has been made to it and the algorithm explored in the paper compared to it.
	Satish et al.’s algorithm used a radix sort that is divided into four steps, these steps are analogues of those in the count sort described above: 
	
\begin{enumerate}
	\item Each block loads and sorts its tile in shared memory using $b$ iterations of $1-bit$ split. Empirically, they found the best performance from using $b = 4$.
	\item Each block writes back the results to Global Memory, including its 2b-entry digit histogram and the sorted data tile.
	\item Conduct a prefix sum over the $p \times 2^b$ histogram table, which stored in column-major order, to compute global digit offsets.
	\item Each thread block copies its elements to their corresponding output position, each rearranges local data on the basis of $D$ bits by applying $D$ 1-bit stream spits operation, this is done using a GPU scan operation. The scan is implemented as a vertical scan operation where each thread in a warp is assigned a set of items to operate on each thread produces a local sum, and then a SIMT operation is performed on these local sums.
\end{enumerate}	

Satish et al. \cite{satish2009designing} stated that the main bandwidth considerations when implementing a parallel radix sort. 
A program making efficient use of available memory bandwidth: 
\begin{enumerate}
	\item Minimizes the number of scatters to global memory.
	\item Maximizing the coherence of scatters. 
	\item Processes 4 elements per thread or 1024 elements per block.
\end{enumerate}	

Given that each block processes $O(t)$ elements, the authors expect that the number of buckets $2^b$ should be at most $O( \sqrt{t} )$, since this is the largest size for which we can expect uniform random keys to (roughly) fill all buckets uniformly. They determine empirically that choosing $b = 4$, which in fact happens to produce exactly $t$ buckets, provides the best balance between these factors and the best overall performance. In a subsequent publication Satish et al. \cite{satish2010fast} identified the bottleneck of the algorithm as the 1-bit split GPU primitive as the bottleneck and found that $65\%$ of time spent by the algorithm is spent performing 1-bit split. Bandyopadhyay et al. \cite{bandyopadhyay2010grs} improved on the work of \cite{satish2009designing} by reordering steps of the algorithm to cut in half the number of reads and writes of non key data, and by removing some random access penalties incurred by \cite{satish2009designing} by moving full data records rather than pointers to records during the search process. The first step of the modified algorithm in \cite{bandyopadhyay2010grs} is to compute a per-tile histogram of the keys. Doing so requires only key values to be read, and then it computes the prefix sum and final position of the elements. The savings of \cite{bandyopadhyay2010grs} lie in only reading keys at step 1, this removes half of the non-key memory reads and writes from the algorithm in \cite{satish2009designing}. Merrill et al. \cite{merrill2011high} added further optimisations to the implementation in \cite{satish2009designing}, in particular they implement ``kernel fusion", ``inter-warp cooperation", ``early exit", and ``multi-scan", they also used an alternative more efficient scan operation introduced in \cite{merrill2009parallel}. In a notable recent development is Zhang et al \cite{zhang2012novel} present a pre-processing operation allowing a fixed rather than dynamically determined memory allocation to the set of buckets and to eliminate data dependence within buckets.  

Kernel fusion is the agglomeration of thread blocks into kernels so as to contain multiple phases of the sort network in each kernel, similar to the work of \cite{baraglia2009sorting}. Inter-warp cooperation refers to calculating local ranks and their use to scatter keys into a pool of local shared memory, then consecutive threads can acquire consecutive keys and scatter them to global device memory with a minimal number of memory transactions compared to the work of \cite{satish2009designing}. Early-exit is a simple optimisation which dictates that if all elements in a bucket have the same value for a given radix then no sort is performed in that bucket at that radix. Finally multi-scan is a complex optimization based on the scan implementation of \cite{merrill2009parallel}. The basic idea of the multi-scan scheme is to generalize the prefix scan primitive in \cite{merrill2009parallel} to calculate multiple dependent scans concurrently in a single pass. Merrill et al. \cite{merrill2011high} implemented multi-scan by encoding partial counts and partial sums within the elements themselves (i.e reserving bits in each element for this purpose). Satish et al. \cite{satish2010fast} expanded on the work of \cite{satish2009designing} by reformulating the algorithm so to allow for greater instruction level parallelism, boosting performance by a factor of $1.6$. Chen et al. \cite{chen2009fast} presented a modified version of the radix sort in \cite{satish2009designing} that is able to effectively handle floats and structs as the original implementation only worked for integer sorting. Chen et al. used an alternate bucketization strategy that works by scattering elements of input into mutually sorted sequences per bucket, and sorting each bucket with a bitonic sorting network. Chen et al. also addressed the loss of efficiency that occurs when nearly ordered sequences are sorted by presenting an alternate addressing scheme for nearly ordered inputs. An earlier hybrid radix-bitonic sort is presented in \cite{govindaraju2006gputerasort}, there as in \cite{chen2009fast} it is used primarily as a bucketization strategy leading into an efficient bitonic sort utilizing GPU primitive hardware operations. Ye et al. \cite{ye2010high} used hardware level synchronization of warps by splitting input into tiles that fit in memory, and then performing a bitonic sort on each tile. Ha et al. \cite{ha2010implicit} compared the radix sort implementation introduced by Satish et al. and Ha et al. and then introduced two revisions, implicit counting and hybrid data representation, to radix sort to improve it. Implicit counting consists of two major components which are the implicit counting number and its associated operations. An implicit counting number encodes three counters in one 32-bit register, each counter is assigned $10$ bits separated by one bit. Using specialized computations that utilized the single 32-bit register allows them to improve running time. The implicit counting function allows the authors to compute the four radix buckets with only a single sweep resulting in this algorithm being twice as efficient as the implicit binary approach of Satish et al.
	
Hybrid data representation improves memory usage. To increase the memory bandwidth efficiency in the global shuffling step, which pairs a piece of the key array and of the index array together in the final array sequence, Ha et al. \cite{ha2010implicit} proposed a hybrid data representation that uses Structure of Arrays (SoA) as the input and Array of Structures (AoS) as the output. The key observation is that though the proposed mapping methods are coalesced, the input of the mapping step still come in fragments, which they refer to as a non-ideal effect. When it happens, the longer data format (i.e. int2, int4) suffers a smaller performance decrease than the shorter ones (int). Therefore, the AoS output data structure significantly reduces the sub-optimal coalesced scattering effect in comparison to SoA. Moreover, the multi-fragments require multiple coalesced shuffle passes which turns out to be costly. They saw the improvement by applying only one pass on the pre-sorting data. Another popular goal for designing hybrid approaches is a decrease in inter-thread communication, for example Kumari et al. \cite{kumari2014parallel} are able to reduce inter-thread communication as compared to odd-even sort om their parallel radix, selection sort hybrid.

\section{Applications and Domain Specific Research}
In this survey we have primarily discussed the fundamental research, techniques and incremental improvements of sorting records of any kind on GPUs. This limits the scope of the survey, but allows for the focused discussion of commonalities that we have presented. However, we would be remiss if we failed to mention 
some of the important work done in optimizing GPU based sorting for application domains. In this section we will briefly mention some of the most interesting work in improving the efficacy of GPU based sorting in specific domains or as a sub-component of an application or system.

Kuang et al. \cite{kuang2009practical} discuss how to use a radix sort for a $KNN$ algorithm in parallel on GPUs. The implementation of the radix sort is divided into four steps, the first three of which are very similar to Satish et al.'s \cite{satish2009designing} version of radix sort, except the fourth steps are different. For the last step of this radix sort, using prefix sum results, each block copies its elements to their corresponding output position. This sorting algorithm can reach many times in performance compared with CPU quicksort. In the final performance test, the sorting phase occupied the largest proportion of the overall computing time, making it the bottleneck in performance of the whole application.

Oliveira et al. \cite{silva2013exploiting} worked on techniques to leverage the space and time coherence occurring in real-time simulations of particle and Newtonian physics. The researchers describe that efficient systems for collision detection play an important part in such simulations. Many of these simulations model the simulated environment as a three dimensional grid. The systems used to detect collisions in this grid environment are bottle-necked by their underlying sorting algorithms. The systems being modeled share the feature that the distribution of particles or bodies in the grid described is rarely uniformly random, and in \cite{silva2013exploiting} two common strategies that leverage uneven particle/body distributions in the grid are described and empirically evaluated. The strategies are described earlier in this survey as they apply to the general problem. 

Finding the most frequently occurring items in real time (coming in from a data stream) has many applications for example in business, and routing. Erra et al. \cite{erra2012frequent} formally state variations of this problem; they describe and analyze two solution techniques aimed at tackling this problem (counting and sorting). Due to the real time requirements of the problem and due to the sensitivity of counting approaches to distribution skew Erra et al. settled on a GPU based sorting solution. The researchers give a performance analysis and an experimental analysis of both approximate and exact algorithms for real-time frequent item identification with a GPU sort as the bottleneck of the procedures \cite{erra2012frequent}. 

Yang et al.'s recent work \cite{yang2012optimized} is focused on improving GPU based parallel sample sort performance on uncommon (at least in GPU sorting literature) distributions such as staircase distributions. As mentioned previously The performance of sample and radix sorts are susceptible to skew in the sorted distributions, in this paper the researchers focus on designing splitters to take advantage of partial ordering found in certain distributions with high skew. The authors offer design considerations and empirical results of several designs taking advantage of the partial ordering found in these distributions.

String or variable length record sorting is an important application in a variety of fields including bioinformatics, web-services, and database applications. Neelima et al. \cite{neelima2014string} give a brief review of the applicable sorting algorithms as they apply to to string sorting and they give a reference parallel quicksort implementation. Davidson et al. \cite{davidson2012efficient} and Drozd et al. \cite{drozd2014efficient} each present competitive modern algorithms aimed at improving the performance of GPU enable sorting of variable length string data. 

Davidson et al.'s parallel merge sort benefits primarily from the speedup inherent in enabling thread-block communication via registers rather than shared memory and are otherwise notable for the fact that their techniques are general for both records with fixed and variable length and achieve the best current sort performance in both cases. Drozd et al.'s parallel radix sort implementation overhead reduces communications overhead by introducing a hybrid splitter strategy which dynamically chooses parallel iterative parallel execution or recursive parallel execution based on conditions present at the time the choice is made.       
Of the two strategies Davidson et al.'s is more general as it applies to strings of any size, however it is possible that Drozd et al's strategy may outperform Davidson et al.'s for small strings; this comparison has not been made in the literature at this time.   

\section{Key Findings}
Our research led to the following general observations: 
\begin{itemize}
	\item Effective parallel sorting algorithms must use the faster access on-chip memory as much and as often as possible as a substitute to global memory operations. All threads must be busy as much of the time as is possible. Efficient hardware level primitives should be used whenever they can be substituted for code that does not use such primitives. Mark Harris’s reduction, scatter, and scan were all hallmarks of the sorts we encountered. 
	\item On the other hand algorithmic improvements that used on-chip memory and made threads work more evenly seemed to be more effective than those that simply encoded sorts as primitive GPU operations.
	\item Communication and synchronization should be done at points specified by the hardware, Ye et al.’s warpsort \cite{ye2010high} is able to achieve the status of fastest comparison based sort for some time using just this approach.
	\item Which GPU primitives (scan and 1-bit scatter in particular) are used made a big difference. Some primitive implementations were simply more efficient than others and some exhibit a greater degree of fine grained parallelism than others as \cite{merrill2009parallel} shows.
	\item Govindaraju et al. \cite{govindaraju2006gputerasort} showed that sorting data too large to fit into memory using a GPU coprocessor while initially thought to be a more significant problem than in memory sort, can be encoded as an external sort with near optimal performance per disk I/O, and the problem then degenerates to an in memory sort.
	\item A combination of radix sort, a bucketization scheme, and a sorting network per SP seems to be the combination that achieves the best results.
	\item Finally, more so than any of the other points above, using on-chip memory and registers as effectively as possible is key to an effective GPU sort.
\end{itemize}	

It appears that the main bottleneck for sorting algorithms on GPUs involves memory access. In particular, the long latency, limited bandwidth, and number of memory contentions (i.e., access conflicts) seem to be major factors that influence sorting time. Much of the work in optimizing sort on GPUs is centred around optimal use of memory resources, and even seemingly small algorithmic reductions in contention or increases in on chip memory utilization net large performance increases. 

\section{Trends in Future Research}
Having reviewed the literature we believe that the following methods are ways to better optimize sorting performance:
\begin{itemize}
	\item Ongoing research aims at improving parallelism of memory access via GPU gather and scatter operations. Nvidia’s Kepler architecture \cite{nvidia_kepler_2012} appears to be moving towards this approach by allowing multiple CPU cores to simultaneously send data to a single GPU.
	\item Key size reduction can be employed to improve throughput for higher bit keys, assuming that the number of elements to be sorted is less than the maximum value of a lower bit representation. 
	\item To reduce memory contention (e.g., cache block conflicts), specific groups of SMs can be dedicated to specific phases of the sorting algorithms, which operate on separate partitions of data. This may not be possible with only one set of data (for particular algorithms), but if we have two or more sets of data that needs to be individually sorted, then we can pipeline them. This is under the assumption that there is enough memory space to efficiently handle such cases. 
\end{itemize}

\section{Conclusion}
It is clear that GPUs have a vast number of applications and uses beyond graphics processing. Many primitive and sophisticated algorithms may benefit from the speed-ups afforded to the commodity availability of SIMD computation afforded by GPUs. We studied the body of work concerning sorting on GPUs. Sorting is a fundamental calculation common to many algorithms and forms a computational bottleneck for many applications. By reviewing and inter-associating their work we hope to have provided a map into this topic, it's contours, and key issues of note for researchers in the area. We predict that future GPUs will continue to improve both in number of processors per device, and in scalability improvements including inter-unit communication. Improvements in handling software pipelines, in globally addressable storage are currently ongoing.

\subsection{Characteristics Of The Literature}
In table \ref{tab:literature_features} we classify the works we reviewed according to features that we think are helpful in guiding a review of the topic, or directing a more narrow review or certain topic features. In deciding which features to include (especially given width limitations), and in communicating the data in the table we were forced to make certain trade-offs. The reader will notice that our table does not include features related to memory use within the GPU or other ``memory utilization" issues; this is because in all or nearly or the work on this topic (with the exception of exclusively theoretical early work) memory utilization of one form or another is critical to the functioning of the algorithm presented. A check mark in our table indicates that a substantial component of the contribution of the article in question relates directly to the highlighted feature. For example many articles present a comparison sort, but contrast it's performance to a radix-based sort presented in an earlier paper, in this case we would not mark the article as a ``Radix Family" article. Regarding the ``Theory" feature, we mark only articles which are primarily or exclusively dedicated to a theoretical analysis of parallel sorting, or develop rather than reiterate such an analysis. Unfortunately many articles do not explicitly state whether the data items sorted during the experiments described in the article involved fixed or variable length data; our assumption in such cases is that the data sorted were fixed length and these articles are marked as ``Fixed Length Sort". Of course it should be clear that sorting algorithms that handle variable length data can also handle fixed length data as a special case, thus we only mark articles in both these categories when fixed length data sorting and variable length sorting are addressed separately and a distinction is made. Most of the sampling and radix based sorts we reviewed us a comparison based sort as a sub-component; articles that make only passing mention or do not describe or develop the comparison based local sorts do not receive a check-mark in our ``Comparison Sort" column. Finally, our ``Misc" column is a catch-all that represents important topics that are simply outside the scope of the features selected these include external sorting, massively distributed or asynchronous sorting, financial analysis of the sort, and other important but uncommon (in the reviewed literature) attributes. 

\section*{Acknowledgment}
We would like to acknowledge the help and hard work of Tony Huynh and Ross Wagner, computer science alumni of UC Irvine 
for their contributions to this survey. We gratefully acknowledge the support of UCCONNECT while preparing this survey. 
Finally, we would like to thank and acknowledge all the researchers whose work was included in our survey. 

\newpage 

\onecolumn
\tiny
\setlength\LTleft{0pt}
\setlength\LTright{0pt}
\begin{longtable}{| c | c | c | c | c | c | c | c | c | c | c |} \hline
    \makecell{Article} & 
    
    \makecell{Comparison \\ Sort} & 
    \makecell{Fixed \\ Length \\ Sort} & 
    \makecell{Variable \\ Length \\ Sort} & 
    \makecell{Primitive \\ Focused} & 
    \makecell{Sampling \\ Family} & 
    \makecell{Radix \\ Family} &
    \makecell{Hybrid \\ Family} &
    \makecell{Theory} &
    \makecell{Domain \\ Specific \\ Sort} &
    \makecell{Misc} \\ [0.5ex] \hline \hline
    
    Bandyopadhyay et al. \cite{bandyopadhyay2013sorting} & 
        \checkmark & \checkmark & \checkmark & \checkmark & \checkmark & \checkmark & \checkmark &  & & \\ \hline
    
    Capannini et al. \cite{capannini2012sorting} & 
        \checkmark & \checkmark &   &   &   & \checkmark & \checkmark &  &  &\\ \hline
        
    Batcher \cite{batcher1968sorting} & 
        \checkmark & \checkmark &   &   &   &   &  & \checkmark & & \\ \hline
        
    Dowd et al. \cite{dowd1989periodic} & 
        \checkmark & \checkmark &   &   &   &   &  & \checkmark & &\\ \hline
        
    Blelloch et al. \cite{blelloch1991comparison} & 
        \checkmark & \checkmark &   & \checkmark & \checkmark & \checkmark & \checkmark &  & & \\ \hline

    Dusseau et al. \cite{dusseau1996fast} & 
        \checkmark & \checkmark &   &   & \checkmark & \checkmark &   & \checkmark & & \\ \hline   

    Cederman et al. \cite{cederman2008practical} & 
        \checkmark & \checkmark & & \checkmark & & & \checkmark & \checkmark & & \\ \hline   
    
    Cederman et al. \cite{cederman2009sorting} & 
        \checkmark & \checkmark &   &  \checkmark &   &   & \checkmark &   & & \\ \hline      
        
    Bandyopadhyay et al. \cite{bandyopadhyay2010grs} & 
        \checkmark & \checkmark & \checkmark &   & \checkmark & \checkmark &   &   & & \\ \hline     
    
    Davidson et al. \cite{davidson2012efficient} & 
        \checkmark & \checkmark & \checkmark &  &  &  & \checkmark &  &  & \\ \hline  
        
    Ye et al. \cite{ye2010high} & 
        \checkmark & \checkmark &   &   &  \checkmark &   & \checkmark &   &  &  \\ \hline 
        
    Dehne et al. \cite{dehne2012deterministic} & 
          & \checkmark &   &   & \checkmark & \checkmark &   &   &  &  \\ \hline  

    Satish et al. \cite{satish2009designing} & 
        \checkmark & \checkmark &   & \checkmark &   & \checkmark &   &   &   & \\ \hline  

    Govindaraju et al. \cite{govindaraju2006gputerasort} & 
        \checkmark &  & \checkmark &  &  & \checkmark & \checkmark &   &  & \checkmark \\ \hline           

    Baraglia et al. \cite{baraglia2009sorting} & 
        \checkmark &   & \checkmark &  &  &  &  &  & \checkmark & \checkmark \\ \hline          

    Manca et al. \cite{manca2015cuda} & 
        \checkmark & \checkmark &  & \checkmark &  &  & \checkmark &  &  &  \\ \hline      
        
    Le Grand et al. \cite{le2007broad} & 
        \checkmark & \checkmark &  &   &  & \checkmark &  &  & \checkmark &  \\ \hline        
        
    Sundar et al. \cite{sundar2013hyksort} & 
        \checkmark & \checkmark &  &  & \checkmark &  & \checkmark &  &   &  \\ \hline         

    Beliakov et al. \cite{beliakov2015parallel} & 
        \checkmark & \checkmark & & \checkmark & \checkmark & & \checkmark & \checkmark  & \checkmark & \checkmark \\ \hline

    Leischner et al. \cite{leischner2010gpu} & 
        \checkmark & \checkmark &  &    & \checkmark &  &  &    &   &   \\ \hline
        
    Ye et al. \cite{ye2014gpumemsort} & 
        \checkmark & \checkmark &  &  & \checkmark &  &  &    &   &   \\ \hline
        
    Chen et al. \cite{chen2009fast} & 
        \checkmark &  & \checkmark &  & \checkmark &  &  \checkmark &    &   &   \\ \hline
    
    Satish et al. \cite{satish2010fast} & 
          & \checkmark &   &  &  & \checkmark &   &    &   & \checkmark  \\ \hline
    
    Merrill et al. \cite{merrill2011high} & 
          & \checkmark &   & \checkmark &  & \checkmark &   &    &   &  \\ \hline
          
    Sun et al. \cite{sun2009count} & 
              & \checkmark &   &   &  &  &  &    &   & \checkmark \\ \hline
              
    Jan et al. \cite{jan2013parallel} & 
              \checkmark & \checkmark &  &  &  &  &  &    &   &   \\ \hline
              
    Ajdari et al. \cite{ajdari2015version} & 
              \checkmark & \checkmark &  &  &  &  &  &    &   &   \\ \hline
              
    Tansic et al. \cite{tanasic2013comparison} & 
              \checkmark & \checkmark &  &  &  &  &  &    &   & \checkmark  \\ \hline
              
    Nvidia \cite{nvidia2010guide} & 
                 &   &  &  &  &  &  & &  & \checkmark    \\ \hline   
              
    Brodtkorb et al. \cite{brodtkorb2013graphics} & 
                &   &  &  &  &  &  &    &   & \checkmark  \\ \hline
                
    Peters et al. \cite{peters2011fast} & \checkmark
                \checkmark &  \checkmark &  &  &  &  &  &    &   &    \\ \hline      
                
    Bandyopadhyay et al. \cite{bandyopadhyay2011sorting} & 
                \checkmark &   & \checkmark &  & \checkmark & \checkmark  &  &    &   &    \\ \hline   
                
    Sengupta et al. \cite{sengupta2007scan} & 
                 &   &  & \checkmark &  &  &  & & \checkmark &    \\ \hline   
                 
    Sengupta et al. \cite{sengupta2008efficient} & 
                 &   &  & \checkmark &  &  &  & & \checkmark &    \\ \hline   
                 
    Keckler et al. \cite{keckler2011gpus} & 
                 &   &  &  &  &  &  & &  & \checkmark    \\ \hline   
          
    White et al. \cite{white2012cuda} & 
                 \checkmark & \checkmark  &  &  &  &  &  & &  & \checkmark    \\ \hline   

    Zhong et al. \cite{zhong2012efficient} & 
                  & &  &  &  &  &  & &  & \checkmark    \\ \hline  
                  
    Greb et al. \cite{greb2006gpu} & 
                   \checkmark &  \checkmark &  &  &  &  &  & &  & \checkmark    \\ \hline  
                   
    Bilardi et al. \cite{bilardi1989adaptive} & 
                   \checkmark &  \checkmark & \checkmark &  &  &  &  & \checkmark &  &      \\ \hline  
                   
    Zachman et al. \cite{zachmann2013adaptive} & 
                   \checkmark &  \checkmark & \checkmark &  &  &  &  & \checkmark &  &      \\ \hline  
                   
    Peters et al. \cite{peters2012novel} & 
                   \checkmark &  \checkmark & \checkmark &  &  &  & \checkmark & \checkmark &   & \\ \hline  

    Kipfer et al. \cite{kipfer2004uberflow} & 
                   \checkmark &  \checkmark &   &  &  &  &  &  &  \checkmark   &     \\ \hline 
                   
    Govindaraju et al. \cite{govindaraju2005fast} & 
                   \checkmark &  \checkmark &   & \checkmark &  &  &  &  &  \checkmark   &     \\ \hline 
                   
    Sintorn et al. \cite{sintorn2008fast} & 
                   \checkmark &  \checkmark &   & \checkmark & \checkmark &  & \checkmark &   &  \checkmark  &     \\ \hline

    Polok et al. \cite{polok2014fast} & 
                     &  \checkmark &   & \checkmark  &  & \checkmark &  &   &    &     \\ \hline                
                     
    Li et al. \cite{shi1992parallel} & 
                     \checkmark &  \checkmark &   &   &  &   &  & \checkmark   &  \checkmark   & \checkmark    \\ \hline  
                     
    Merrill et al. \cite{merrill2009parallel} & 
                 & \checkmark  &  & \checkmark &  &  \checkmark &  & &  &    \\ \hline   
                 
    Zhang et al. \cite{zhang2012novel} & 
                 & \checkmark  &  &   & \checkmark &  \checkmark & \checkmark & &  &    \\ \hline
                 
    Ha et al. \cite{ha2010implicit} & 
                 & \checkmark  &  &   &   &  \checkmark &   & &  &    \\ \hline
                 
    Kumari et al. \cite{kumari2014parallel} &
               & \checkmark  &  &   &  \checkmark  &  \checkmark & \checkmark & &  &    \\ \hline
               
    Kuang et al. \cite{kuang2009practical}& 
                 & \checkmark  &  &   &   &  \checkmark &  & & \checkmark &    \\ \hline
                 
    Oliveira et al. \cite{silva2013exploiting} & 
                \checkmark  & \checkmark  &  &   &   &  \checkmark &  & \checkmark & \checkmark &    \\ \hline
                
    Erra et al. \cite{erra2012frequent} & 
                  & \checkmark  &  &   &   &  \checkmark &  &  & \checkmark &    \\ \hline
                  
    Yang et al. \cite{yang2012optimized} & 
                  \checkmark & \checkmark  &  &   & \checkmark  &    &  &  &   & \checkmark    \\ \hline
                  
    Neelima et al. \cite{neelima2014string} & 
                  \checkmark &    & \checkmark &   &   &    &  &  &   \checkmark &     \\ \hline
                  
    Drozd et al. \cite{drozd2014efficient} & 
                    &   & \checkmark &   &   &  \checkmark  &  &  &   \checkmark &     \\ \hline
                    
    Nvidia \cite{nvidia_kepler_2012} & 
                 &   &  &  &  &  &  & &  & \checkmark    \\ \hline   
                
\caption{Features of the Literature} 
\label{tab:literature_features}
\end{longtable}
\normalsize

\twocolumn

\vfill

\bibliographystyle{IEEEtranN}
\bibliography{}


\end{document}